\documentclass[%
pra,
 amsmath,amssymb,
 reprint,showkeys, %
superscriptaddress,longbibliography
]{revtex4-1}
\usepackage{braket}
\usepackage{hyperref}
\usepackage{ulem}
\usepackage{mathtools}
\usepackage{graphicx}
\usepackage{dcolumn}
\usepackage[dvipsnames]{xcolor}
\usepackage{bm}
\graphicspath{{fig/}}
\newcommand{\affilITMO}{ITMO University, Birzhevaya liniya 14, 199034 St.-Petersburg, Russia}

\newcommand{\affilNone}{Rehovot 7630567, Israel }
\usepackage{physics}
\usepackage{appendix}

\renewcommand{\Im}{\operatorname{Im}}

\newcommand{\te}[1]{\widehat{{#1}}}

\newcommand{\rmi}{{\rm i}}

\newcommand {\e}{{\rm e}}

\begin{document}

\preprint{AIP/123-QED}

\title[]{
Chirality-driven delocalization in disordered waveguide-coupled quantum  arrays
}

\author{Gleb Fedorovich}
  \email{gleb.fedorovich@metalab.ifmo.ru}
 \affiliation{\affilITMO}

\author{Danil Kornovan}%

\affiliation{\affilITMO 
}%

\author{Alexander Poddubny}
\affiliation{\affilNone}

\author{Mihail Petrov}
\affiliation{\affilITMO
}%


\begin{abstract}
We study theoretically the competition between directional asymmetric coupling and disorder in a one-dimensional array of quantum emitters chirally  coupled  through a waveguide mode.   Our  calculation reveals highly nontrivial phase diagram for the eigenstates spatial profile, nonmonotonously depending on the  disorder and directionality strength. The increase of the  coupling asymmetry drives the transition from Anderson localization in the bulk through delocalized states to chirality-induced localization  at the array edge. Counterintuitively, this transition is not smeared by  strong disorder but becomes sharper instead.   Our  findings could be important for the rapidly developing field of the waveguide quantum electrodynamics, where the chiral interactions and disorder play crucial roles.
\end{abstract}

\maketitle

 \section{Introduction.} Localization of waves and particles in disordered media remains one of the key universal concepts in modern physics \cite{Lagendijk2009} starting from the first theoretical prediction by Anderson \cite{Anderson1958}.  One-dimensional systems are especially remarkable since, according to the classical scaling theory \cite{Abrahams1979}, all of the states are localized for an arbitrarily weak disorder. However, the situation changes drastically in non-Hermitian disordered quantum systems, where one can observe  both localized and delocalized states \cite{Hatano1996,Brouwer1997,Brouwer1998,Hebert2011,Feinberg1999}. In this regard, one of the most interesting platforms  is offered by waveguide quantum electrodynamics (WQED)~\cite{Chang2018,Sheremet}, studying interactions of localized quantum emitters with photons propagating in a one-dimensional waveguide. Such a system  is inherently  strongly non-Hermitian due to the presence of radiative  losses and also features long-range light-induced couplings, that have recently been predicted to suppress localization~\cite{Haakh2016}.   
 Photon-photon interactions driven by  anharmonicity of the emitter Hamiltonians can enable quantum chaos \cite{Zhong2021chaos}, and many-body localization~\cite{Fayard2021}. The effects of disorder have been also extensively studied in an alternative non-Hermitian system based on semiconductor polaritonic lattices~\cite{Malpuech1999,Kosobukin2003,Kosobukin2007}  {and complex systems with loss and gain \cite{Lubatsch2010}}.

\begin{figure}[t]
\includegraphics[width=1\columnwidth]{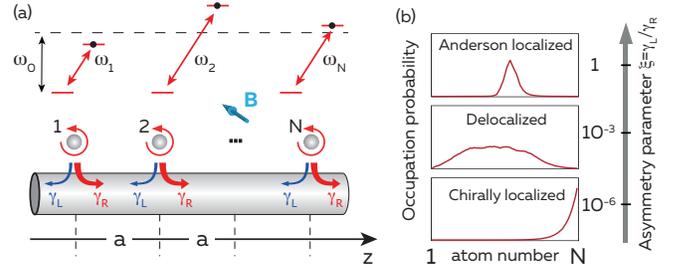}
\caption{(a) The geometry of an array of regularly spaced quantum emitters separated by a distance $a$ and directionally coupled through a waveguiding mode. (b) Localized and delocalized eigenstates depending on the coupling directionality parameter $\xi\equiv \gamma_L/\gamma_R$ calculated for $\delta=0.05$ and $N=100$, having the eigenfrequencies $(\omega-\omega_0)/\gamma_0=0.48,0.03,0.01$, respectively.}\label{fig:1_Main}
\end{figure}

The situation becomes even more interesting in the regime of chiral quantum optics~ \cite{Lodahl2017}, when  a {constant} magnetic field  is applied transverse to the waveguide {introduces artificial ``chirality" to the system and} makes light-induced  couplings between the atoms partially unidirectional [see Fig.~\ref{fig:1_Main}a]. {The directional  coupling appears due to  polarization dependent  waveguide mode excitation. } This destroys the internal symmetry of the problem and also suppresses the quantum interference effects responsible for localization. 
However, despite recent numerical studies of the  photon transmission and reflection  through chiral disordered atomic arrays~Ref.~\cite{Mirza2017,Mirza2018,Jen2020a} and the recent interest to non-Hermitian skin effects~\cite{Okuma2020, Bergholtz2021, Okuma2022}, as well as scaling theory of localization in chiral non-Hermitian systems~ \cite{Kawabata2020}, the fundamental problem of localization in non-Hermitian disordered systems with directional couplings is still open.

Here, we study theoretically localization and delocalization of a single excitation in the  array of atoms with fluctuating frequencies, depending on the fluctuation strength and the directionality of the atom-waveguide mode coupling.  We reveal a delicate competition between the chirality and disorder strength. {The origin of the competition is straightforward:  disorder tends to localize the states in the bulk area, while chirality tends to localize the states at the edge of the system. 
We show that,} counterintuitively, the effect of a chiral coupling is not universal and it can either localize or delocalize eigenstates depending on the disorder strength, as is schematically illustrated in Fig.~\ref{fig:1_Main}b. 
The advantage of the considered atomic chiral setup is its high coherence and tunability  by external magnetic field\cite{Corzo2016}. However, our theoretical results are quite general and apply both to quantum and classical chiral systems. For example, topological photonic structures~\cite{Lu2014}, where the unidirectional propagation of protected edge states is one of the central scenarios, attract now a lot of interest \cite{Barik2020,JalaliMehrabad2019,JalaliMehrabad2020}.


\section{Theoretical framework} 

\subsection{General description of a chiral disordered array} 
We start with the consideration of  a  one-dimensional (1D) array of $N$ two-level quantum emitters  placed at the coordinates $z_n\equiv nd, n=1,2...$, and coupled through a single guided mode, which is schematically shown in Fig.~\ref{fig:1_Main} (a). In the case of a finite system, the effective Hamiltonian can be represented as $\widehat{H} = \widehat{H}_0+ \widehat V$   with \cite{Asenjo-Garcia2017a}:
\begin{equation}
\label{eq:H}
\widehat{H}_0= {\hbar}\sum\limits_{m=1}^{N}\left(\omega_m- i\dfrac{\gamma_0}{2} \right)\widehat\sigma_{m}^+\widehat\sigma_{m}^-, \quad 
\widehat V= \hbar \sum\limits_{{}^{m,n=1}_{m \ne n}}^{N}g_{n,m}\widehat\sigma_{n}^+\widehat\sigma_{m}^-,
\end{equation}
here $\omega_m\equiv \omega_0+\Delta\omega_m$, and $\gamma_m$ are the transition frequency and radiative emission rate of the $m$-th emitter, respectively, and $g_{n,m}$ are the emitter-emitter coupling constants. 
We  focus on the diagonal disorder due  to the fluctuations of  transition frequencies of the $m$-th emitter so that the fluctuations $\Delta \omega_m$ are  normally distributed random numbers  with standard deviation equal to $\delta\cdot\gamma_0$ in the absence of correlations between the emitters. The   radiative emission rates are assumed to be constant for all emitters,  $\gamma_m=\gamma_0$. { The proposed theoretical model can potentially find an experimental realization, for instance, in a cold-atomic array localized in the vicinity of a nanofiber in a periodic optical potential \cite{Nayak2018} with random fluctuations or a nanofiber with corrugated surfaces. The fluctuating stable atom-fiber distance will provide random Lamb shift in the energy of atomic transitions. Alternatively, one may suggest superconducting circuit \cite{Besedin2021} with random inharmonicity, which will also contribute to random fluctuations of the transition energy of artificial atoms.  }  

The interemitter coupling constants $g_{n,m}$  can be expressed through the electromagnetic Green's function \cite{Gruner1996,Asenjo-Garcia2017a}. They depend on the polarization properties of both the guided mode, and the transition dipole moments, and take the form $g_{n,m} = -i\gamma_{R} e^{i\varphi_{nm}}\ \text{for}\ m>n,$ and $g_{n,m} = -i\gamma_{L} e^{i\varphi_{nm}}\ \text{for}\ m<n$,
where  $\gamma_R=\gamma_0/(1+\xi)$ and $\gamma_L=\xi\gamma_R$ are emission rates to the right and left directions,  correspondingly, and parameter $\varphi_{nm}=k_0|z_n-z_m|$ is the phase due to propagation of a photon between the emitters $n$ and $m$. The parameter $\xi, 0\le\xi\le 1,$ characterizes the degree of asymmetry.


Here, we focus on singly excited {quasistationary} states of the emitter array $|\psi_m\rangle=\sum_{n=1}^Nc_{nm}\sigma_n^+|0\rangle$, which  are  collective polaritonic states formed due to the long-range coupling of  emitters through the guided mode \cite{Ivchenko1994, Vladimirova1998,Angelatos2016, Kornovan2016}. Their eigenfrequencies $\Omega_k$ in a finite structure  are complex valued due to the radiative decay rate, and can be found from the following  Schr\"odinger equation:
\begin{equation}
\te H \ket{\psi_m}=\hbar \Omega_m \ket{\psi_m},\quad \Omega_m=\omega_m-i\gamma_m/2.\label{eq:eigenmodes}
\end{equation}
More details on the eigenstates of regular infinite and finite periodic structures are provided in Appendices  \ref{App A} and  \ref{App B}, respectively.
The following analysis of the effects of disorder  relies on the properties of eigenfunctions $|\psi_{m}\rangle$ of the equation above. However, the localization effects can also be manifested in the optical response of the system, for example, the transmission coefficient,  and the next subsection will cover this aspect. 

\subsection{Localization length estimated from the transmission spectra}\label{appendix:loclength}

Another subject we want to cover in this section is how to extract the localization length from the transmission coefficient through the structure. 

In principle, propagation and localization of waves in one-dimensional disordered  structures should be described by a general phase formalism \cite{lifshitz1988} that has been successfully applied to photonic structures, see Ref.~\cite{Poddubny2012} and references therein. However, generalization of the phase formalism for the case of directional coupling is a separate task that lies out of the scope of the current manuscript. Instead, we resort here to a more simplified semi-phenomenological approach that ignores interference of waves reflected from different atoms but still captures the essence of light localization away from the resonance frequency $\omega_0$.
Specifically, the reflection coefficient of light from the $m$-th atom can be presented as:
\begin{align}\label{eq:rm}
r_m &=\frac{\sqrt{\gamma_L\gamma_R}}{\omega_m-\omega-\rmi \gamma_0/2}\approx 
\langle r\rangle +\delta r_m\:,\\
\langle r\rangle&=\frac{\sqrt{\gamma_L\gamma_R}}{\omega_0-\omega-\rmi \gamma_0/2}\:,\\
\delta r_m&=-\frac{\sqrt{\gamma_L\gamma_R}}{(\omega_0-\omega-\rmi \gamma_0/2)^2}\delta \omega_m\:,
\end{align}
where $\delta\omega_m\equiv \omega_m-\omega_0$ is the frequency fluctuation, $\gamma_R=\gamma_0/(1+\xi)$ and $\gamma_L=\xi\gamma_R$. Here, $\langle r\rangle$ is the coherent part of the reflection coefficient, responsible for the formation of the polaritonic band gap in the ordered structure. On the other hand, $\delta r_m$ is the disorder-induced reflection, zero on average, but responsible for wave localization. In writing Eq.~\eqref{eq:rm} we have assume that the frequency is far enough from atomic resonance so that it is  sufficient to take into account only one term in the Taylor expansion in powers of frequency fluctuations $\delta\omega_m$.   Our next crucial simplification, assuming strong uncorrelated disorder, is the independent transmission of waves through different atoms, without taking into account multiple reflections:
\begin{equation}\label{eq:approx}
T_{N}=T_{N-1}(1- |\delta r_N^2|)\:.
\end{equation}
Equation~\eqref{eq:approx} tells that the probability of the wave to pass through $N$ atoms is given by the probability of light to pass through $N-1$ atoms times the probability of not being scattered  by the disorder at the $N$-th atom. We stress that Eq.~\eqref{eq:approx} does not take into account the coherent part of the reflection coefficient $\langle r\rangle$ that does not contribute to wave localization away from the band gap and just renormalizes the wave dispersion law.

Given that $\langle \delta r_N^2\rangle\ll 1$ we find from Eq.~\eqref{eq:approx} the decay law for the transmission coefficient
\begin{align}
\ln T_N\propto -\frac{N}{L_{\rm loc}}\label{eq:Lloc0}\:,\\ \frac1{L_{\rm loc}}=\langle |\delta r_m|^2\rangle\:,\label{eq:Lloc1}
\end{align}
where $L_{\rm loc}$ is the length of extinction (being equivalent to the localization length in the case of a one-dimensional system) and the angular brackets denote averaging over the disorder.
Calculating the average $\langle |\delta r_m|^2\rangle$  we obtain the formula that estimates $L_{\rm loc}$ provided that $|\omega-\omega_0|\gg \gamma_0$:
\begin{equation}\label{eq:Lloc_final}
L_{\rm loc}(\omega,\xi)=\frac{(1+\xi)^2(\omega-\omega_0)^4}{\xi \gamma_0^4 \delta^2 }\:.
\end{equation}  
More formal equivalent derivation of Eq.~\eqref{eq:Lloc_final} for the localization length in case of symmetric coupling,   $\xi=1$, based on the phase formalism and the Fokker-Planck equation is presented in Ref.~\cite{Poddubny2012}.

As this equation is only a rough analytical estimate of the localization length, one also needs to calculate a precise numerical value of the transmission coefficient, which can be done using the transfer matrix approach, which is widely known, and theoretical details of which are covered in Appendix \ref{AppB}.

\begin{figure}[t]
\includegraphics[width=0.9\columnwidth]{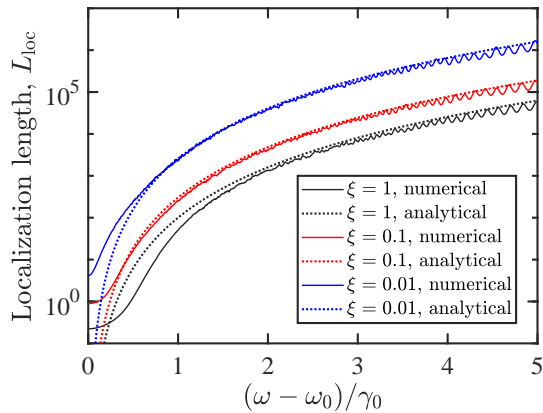}
\caption{Frequency dependence of the localization length calculated numerically from Eq.~\eqref{eq:Lloc0} and analytically from
Eq.~\eqref{eq:Lloc_final} for three different values of the directionality parameter $\xi$, fixed disorder strength $\delta=0.2\gamma_0$, and period that implies $\phi_{n, n+1} = k_0 \Delta z = \pi/2$. The array length was chosen to be $N=1000$, and the averaging has been done over $500$ disorder realizations. }\label{fig:S2}
\end{figure}

Fig.~\ref{fig:S2} presents the comparison of the localization length, calculated numerically by averaging  the logarithm of the transmission coefficient over the disorder, following
Eq.~\eqref{eq:Lloc0}, and analytically, following Eq.~\eqref{eq:Lloc_final}. The analytical and numerical results are in a good quantitative agreement, especially in an expected region $\omega-\omega \gg \gamma_0$ far enough from the bandgap. By this we confirm the qualitative validity of Eq.~\eqref{eq:Lloc_final}, which we will use in the next section when discussing the results.

\section{Numerical results and discussions.} 

\begin{figure*}[!]
\includegraphics[width=\textwidth]{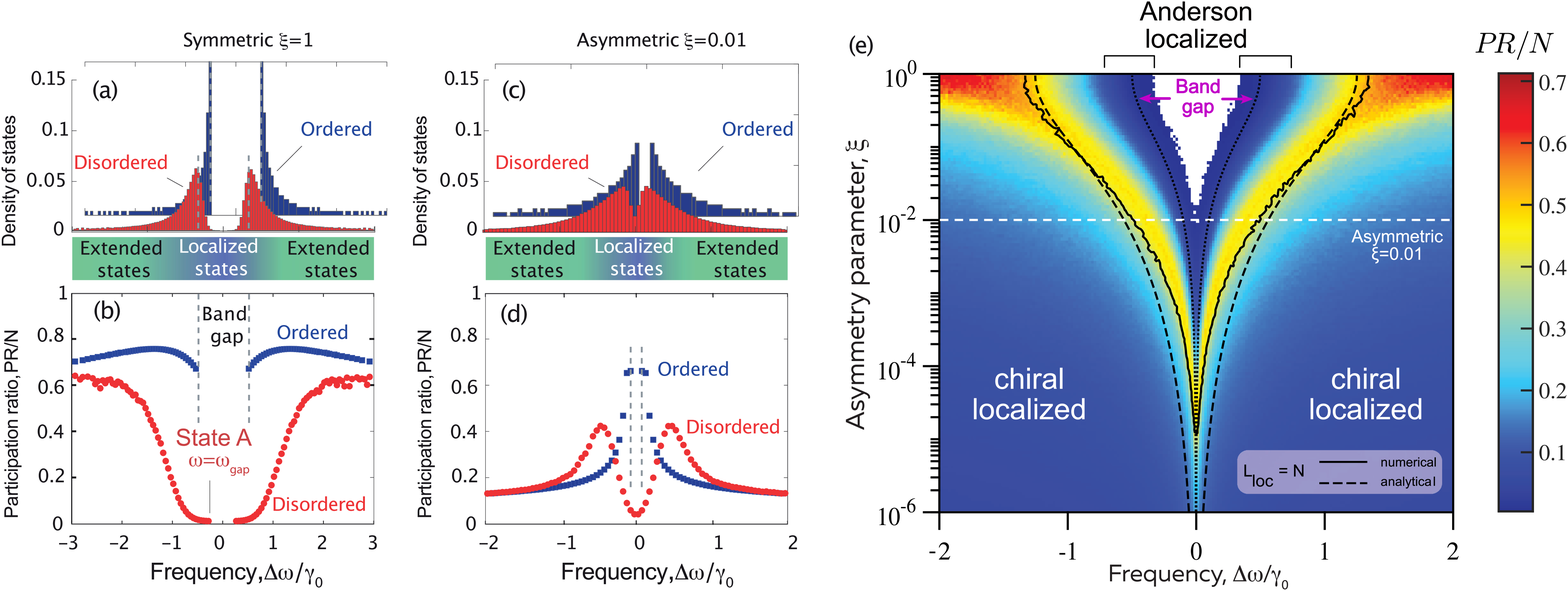}
\caption{{(a--d)} Density of states (a,c) and normalized participation ratios $PR/N$ (b,d)  for  the ordered ($\delta=0$,    blue color) and disordered ($\delta=0.1$, red color) system with symmetric atom-waveguide coupling (a,b) and directional (c,d) coupling with $\xi=0.01$. (e) Normalized participation ratio   of eigenstates  for a fixed disorder amplitude $\delta=0.1$ as a function of the frequency detuning $\Delta \omega$ and asymmetry parameter $\xi$. Blue and red region correspond to localized and delocalized states, respectively. Dotted black curve illustrates the band gap in ordered array, that closes for smaller $\xi$. Solid and dashed black curves show the boundary between localized and extended states $L_{\rm loc}=N$, where the localization length $L_{\rm loc}$ is calculated numerically and analytically. Horizontal dashed line indicates the value $\xi=0.01$ corresponding to panels (c,d). The simulation parameters for (a)--(d) are $N=400$ and $\varphi={\pi}/{2}$, for (e) they are $\delta = 0.1,\ N = 1000$. The results were obtained after averaging over  1000, for (a)--(d), and 100, for (e), random realizations. }\label{fig:4_DOS}
\end{figure*}

We characterize the spectrum of a finite disordered structure with the density-of-states (DOS) function. The spatial distribution of the eigenstates is described by the participation ratio parameter ($PR$) \cite{VanTiggelen1999}, that quantifies the effective  number of the occupied sites by a single excitation, and reads
$PR_m=\left(\sum_{i=1}^{N}\left|c_{im}\right|^{2}\right)^2/\sum_{i=1}^{N}\left|c_{im}\right|^{4}
$
for the $m$-th state. Since the   polaritonic eigenmodes of the ordered periodic array are just  the delocalized Bloch waves, the excitation occupies almost all of the lattice sites and $PR\sim N$.  On the other hand, for a localized eigenstate we expect smaller values of  $PR\sim 1$, independent of $N$. 

The DOS profiles shown in Fig.~\ref{fig:4_DOS}(a) have a typical two-peak structure that can be understood from the polariton dispersion in a periodic array described in Appendix \ref{App A}. The DOS  manifests a band gap around the  frequency $\omega_0$ resulting from the avoided crossing of the light line  with the atomic resonance \cite{Sheremet}. The band gap width is equal to $\gamma_0$ in case of $\varphi=\pi/2, \xi = 1$\cite{Vladimirova1998}, see also  Fig.~\ref{fig:S1}. The DOS function has  van Hove singularities at the gap edges typical for one-dimensional systems. As expected, disorder leads to the smearing of the band edges, and formation of the Urbach tails \cite{VanTiggelen1999}, where the states are strongly localized as can be seen from   Fig.~\ref{fig:4_DOS}(b).  {With the increase of the asymmetry (smaller $\xi$) the polaritonic band gap gets more narrow as  can be directly  seen from  the comparison of  Fig.~\ref{fig:4_DOS}(a,b) with Fig.~\ref{fig:4_DOS}(c,d). For small asymmetry parameters the  band gap width becomes comparable to the energy of Urbach tails, and there appears non-zero density of states in the band gap center with a relatively small value of $PR$ in Fig.~\ref{fig:4_DOS}(d). }	
In the symmetric case $\xi=1$, localization length $L_{\rm loc}$ increases fast when the frequency is detuned from the atomic resonance. For $|\omega-\omega_0|\equiv |\Delta\omega|\gg \gamma_{L,R}$ the localization length can be approximately estimated from the expression we derived in the previous section: 
${L_{\rm loc}(\omega,\xi)={(1+\xi)^2(\omega-\omega_0)^4}/{\xi \gamma_0^4 \delta^2 }}$. 
 In order to distinguish between the localized and extended eigenstates for a finite array it is instructive to compare the localization length with the array size, 
$L_{\rm loc}(\omega,\xi)=N$, 
  because if the array is shorter than the localization length  then the eigenstate is spread over all atoms. {The obtained frequency dependence $L_{\rm loc}(\omega,\xi)$   is shown by the black dashed curve in Fig.~\ref{fig:4_DOS}(e). 
It is in qualitative agreement with the numerical solution of $L_{\rm loc}(\omega,\xi)=N$ where the localization length has been  extracted from the disorder-averaged logarithm of the numerically calculated transmission coefficient through a finite array as $1/L_{\rm loc}=-\langle \ln |t_N^2| \rangle/N$ (solid curve).} Specifically, for $\xi=1$  the central spectral region in Figs.~\ref{fig:4_DOS}(a,b,e) with frequencies $|\Delta\omega/\gamma_0|\lesssim 0.4$ corresponds to the band gap with no eigenstates (white color in the panel (e)), and it is surrounded by a region of Anderson-localized states with $0.4\lesssim |\Delta\omega/\gamma_0|\lesssim 0.7$ (blue color). For even larger detunings $\Delta\omega$ the
localization length exceeds the array size and the states become extended (red color in panel e).

The profile of eigenstates changes dramatically for an asymmetric array with $\xi<1$. The effect of coupling asymmetry is twofold. First, the smaller the $\xi$ the  narrower the spectral region of localized states and the narrower the band gap in the ordered structure (see black curves in Fig.~\ref{fig:4_DOS}e). This is the consequence of the suppression of back reflections and Anderson localization in the strongly chiral setup with $\xi\ll 1$.

Strong asymmetry of the interaction destroys both the polaritonic bands and  Anderson localization in the bulk of the system. However, one can consider an extreme case of $\xi\to0$, when all of the states    are squeezed to the right edge of the system due to chiral localization  (outer blue region in Fig.~\ref{fig:4_DOS}(e) that corresponds to the chiral localization). Such a chiral localization can be also seen as a direct manifestation of a so-called non-Hermitian skin effect \cite{Bergholtz2021,Okuma2022}.
Indeed, for small $\xi$ all of the modes are almost  completely degenerate with {$\Omega_k=\omega_0-\rmi\gamma_0/2$} and are strongly localized at the right edge of the chain even in the absence of disorder.
  This is explained by the fact that each emitter radiates to the left  weaker than to the right in the asymmetric coupling case. In the limit $\xi\to0$  only one non-trivial state survives, {$\ket{N}\equiv \sigma_N^+|0\rangle$}, and it is localized at only a single atom {at the edge of the chain} (see Appendix \ref{appendix:chiral_disorder} for details). 
  When disorder is introduced  the spectral  degeneracy is lifted  and $N$ non-degenerate eigenmodes become smeared over a few sites close to the edge of the system  (see Fig.~\ref{fig:fig_apx_1} in Appendix \ref{appendix:chiral_disorder}) with inverse localization length $1/L_{\text{loc}}$ having a logarithmic dependence on the disorder amplitude as can be seen from Fig.~\ref{fig:fig_apx_1}. In order to obtain this figure, we pick the state with the largest $PR$ for each realization of disorder, average it over multiple  realizations, and fit it with the exponential function  for  the atoms close to the right edge of the chain.
    
\begin{figure*}[t]
\includegraphics[width=0.9\textwidth]{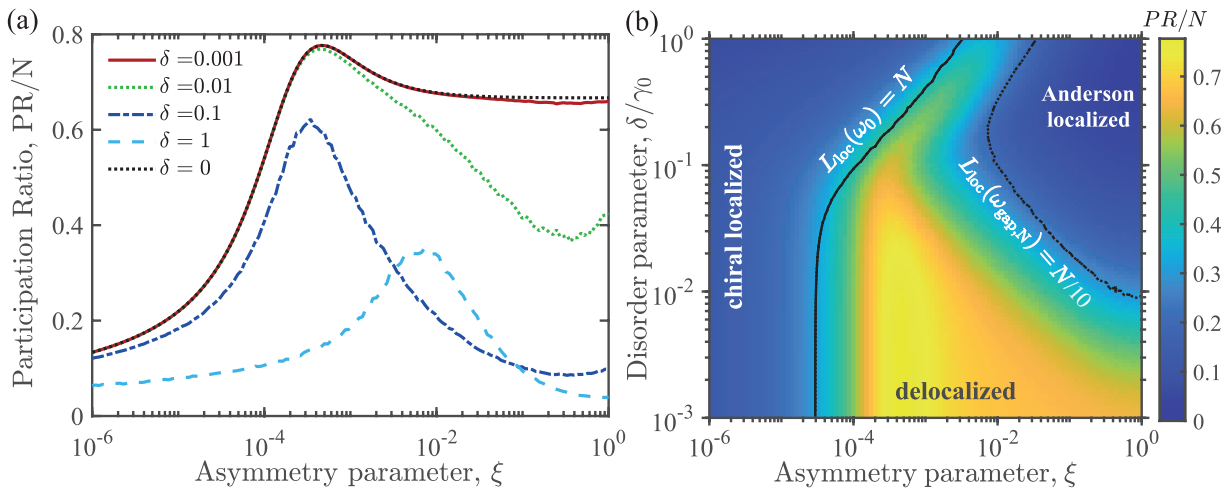}
\caption{(a) Normalized participation ratio for the state A closest to the resonance (see Fig.~\ref{fig:4_DOS}a) calculated depending on the asymmetry parameter $\xi$ for several disorder  strengths $\delta$. (b) False color map of the participation ratio for the State A ($\omega_{\text{gap}}$) depending on both $\xi$ and $\delta$. Black solid, and dashed curves have been extracted from  disorder-averaged localization length defined through a transmittance, the details are explained in the text.
 The number of random realizations used for averaging in (a)--(b) is  500, calculation has been performed for $N=100$ emitters.} \label{fig:5_PhaseMaP}
\end{figure*}

  Finally,  the most striking effect is observed in the transition region for a moderate value of the asymmetry parameter $\xi$. In this case, for a fixed spectral detuning, e.g. $\Delta \omega=0.5\gamma_0$ , the diagram in Fig.~\ref{fig:4_DOS}(e) indicates the appearance of delocalization (at $\xi\approx 0.1$, marked with a dashed white line) on the way of gradual transition from Anderson disorder-induced localization ($\xi\to 1$) to a chiral localization ($\xi \to 0$). {Appearance of the region with a large participation number indicates that chirality suppresses the effect of disorder and states become extended at the scale of the array size. }Since  the results in Fig.~\ref{fig:4_DOS}(e) have been  obtained after averaging the $PR$ of states in a finite energy range, the contributions from  localized and delocalized states could potentially be mixed and affect the average $PR$. In order to verify that this is not the case, it is instructive to follow directly the evolution of individual eigenstates with the increase of asymmetry, which   has been done in Fig.~\ref{fig:5_PhaseMaP}. 
   Specifically, for each disorder realization we select one  eigenstate  that is spectrally closest to the resonance frequency $\omega_0$, see  label A in Fig.~\ref{fig:4_DOS}(b).  Next, we focus on this state and  analyze
   in Fig.~\ref{fig:5_PhaseMaP} the  transition from symmetric to chiral coupling for different values of the disorder amplitude $\delta$. Figure~\ref{fig:5_PhaseMaP}(b) shows the dependence of the $PR$ on  both $\delta$ and $\xi$ and Fig.~\ref{fig:5_PhaseMaP} (a) presents the slices of this dependence for several characteristic disorder strength values $\delta$. The calculation demonstrates that  for symmetric case and small disorder, the eigenstate extends over the whole system. With the increase of disorder amplitude $\delta$, the closest-to-band gap center state occupies a small part of the array  and $PR/N\lesssim 0.1$, which is a sign of Anderson localization in the bulk of the system. 

For an extremely strong asymmetry ($\xi \to 0$) the state naturally becomes localized at the edge of the system due to the directional interaction. {The boundary of this localized-states region can be estimated from  $L_{\text{loc}}(\omega_0,\xi)=N$  
with localization length evaluated numerically from the transmission coefficient through a finite array at the transition frequency, which also corresponds to the closure of the bandgap due to disorder. The corresponding boundary, extracted from the extinction spectra, is shown by a black curve in Fig.~\ref{fig:5_PhaseMaP}(b) and agrees well with the result of the calculation of the participation ratio.}
However,  for moderate values of asymmetry parameter $\xi \gtrsim 10^{-4}$  and relatively weak disorder $\delta\lesssim 0.1$, there exists a transition region, shown by yellow colors in Fig.~\ref{fig:5_PhaseMaP}(b), where the states are extended, and occupy a significant  part of the array. In this transition region the interaction asymmetry leading to edge localization competes with the disorder, which tries to localize the state in the bulk of the system. 
Thus, if we fix disorder strength, the transition from the Anderson localization at $\xi=1$, to a chiral edge localization at $\xi\to 0$ is indeed nonmonotonous, and occurs through extended states.  Counterintuitively, when the disorder strength increases { from $\delta=10^{-3}$ to $\delta=0.1$}, this transition   becomes sharper and shifts, i.e. it occurs in a narrower range of a parameter $\xi$ and for smaller values of $\xi$. {For an even stronger disorder, $\delta>0.1$, the values of $\xi$ corresponding to the transition begin to increase.} {The second (right) boundary between the Anderson localization, and delocalization can be qualitatively found if one equates the localization length (defined through transmittance $T(\omega)$) at the frequency of the state closest to the bandgap for absent disorder in a finite system to, for example, $10\%$ of the system size: $L_{\text{loc}}(\omega_{\text{gap},N})=N/10$. } In case of Fig.~\ref{fig:5_PhaseMaP}(b) this second boundary is shown in dashed black, and it separates the states that are Anderson localized due to disorder from the yellow transition region. {Characteristic disorder-localized, extended and chiral-localized eigenstates are also shown in Fig.~\ref{fig:1_Main}(b).}
        
\begin{figure}[t]
\centering\includegraphics[width=0.5\textwidth]{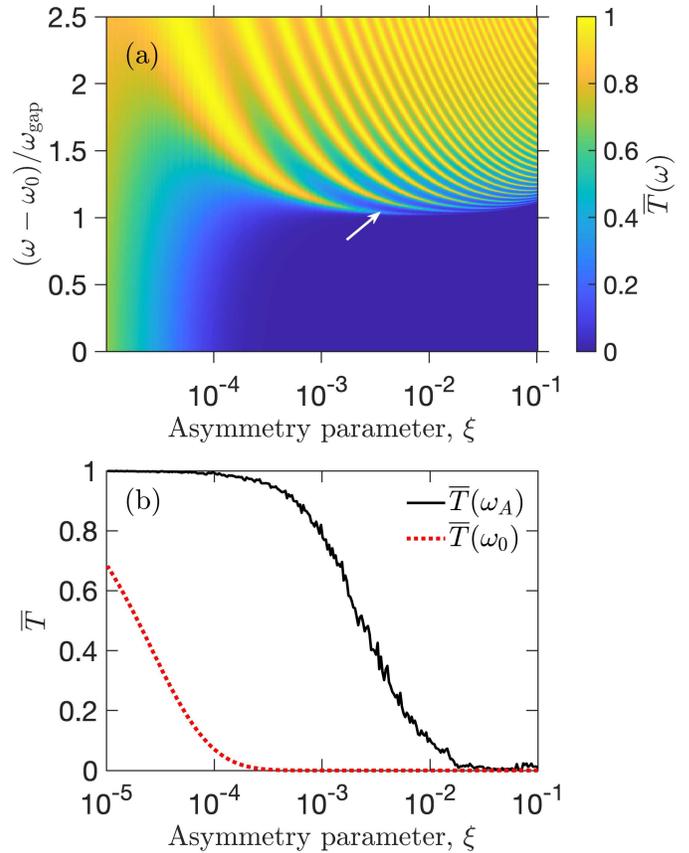}
\caption{(a) Dependence of the transmission spectra on the asymmetry parameter $\xi$. Calculated for a disorder  parameter $\delta=10^{-2}$ .
(b) Transmission coefficient calculated at the resonance frequency and at the frequency of the closest to atomic resonance $\omega_0$ peak, shown by an arrow in the panel (a).
}\label{fig:Tmap}
\end{figure}

The transition from Anderson localization to delocalization to chiral localization can be directly detected in transmission spectrum. To this end we have plotted in 
Fig.~\ref{fig:Tmap}(a) the disorder-averaged transmission spectra $\overline T \equiv \exp[\langle\ln T\rangle]$ for a fixed disorder strength $\delta$. The averaging of transmission logarithm $\langle\ln T\rangle$ has been performed over $N_{\rm av}=60$ disorder realizations. The frequency axis has been normalized to the gap halfwidth in the ordered system $\omega_{\rm gap}$.

 The transmission peaks correspond to the eigenmodes of the finite array and the phase diagram Fig.~\ref{fig:5_PhaseMaP}b   can be reproduced  by tracing the peak dependence on the asymmetry and disorder parameters. In order to illustrate this, we plot in Fig.~\ref{fig:Tmap}(b) the asymmetry dependence of the  transmission coefficient at the resonance, $\overline T(\omega_0)$ and the transmission at  the frequency of the eigenstate closest to the gap center,
$\overline T(\omega_A)$, shown by an arrow in panel (a). Three qualitatively different ranges of the asymmetry parameter can be distinguished.
For $\xi\gtrsim 10^{-2}$ the transmission is suppressed both at the $\omega_A$ frequency and at the resonance frequency. This corresponds to the regime of Anderson localization.
In the intermediate range,  $10^{-4}\lesssim \xi\lesssim 10^{-2}$ the transmission  coefficient $\overline T(\omega_A)$ starts to increase (solid black curve in Fig.~\ref{fig:Tmap}b). This reflects quenching of disorder and delocalization  of the corresponding eigenstate. Finally, when the asymmetry becomes even stronger,  $\xi\lesssim 10^{-4}$, the transmission  coefficient $\overline T(\omega_0)$  also becomes large  (dotted red curve in Fig.~\ref{fig:Tmap}b). This means that the whole array becomes transparent and backscattering is suppressed even at the resonance. Thus, the structure is in the fully chiral regime (when all eigenstates are localized due to strong interaction asymmetry). An analysis of  such transmission maps for different values of disorder strength $\delta$ has allowed us to obtain the black curves $L_{\rm loc}(\xi)\equiv -1/\langle \ln T\rangle$ in Fig.~\ref{fig:4_DOS}b and, thus, to independently reproduce the phase diagram previously found from the study of the spatial profile of the eigenstates.

Taking into account disorder in radiative emission rate of each atom $\gamma_n$ will provide ``non-diagonal" disorder \cite{Biswas2000} which may enable new intriguing effects in the system such as Dyson singularity \cite{Dyson1953, Kozlov1998, Petrov2015}. At the same time, introducing  correlated disorder \cite{Croy2011} will make the disorder contribution in  the chirality-disorder competition   more pronounced  but keeping the physical picture the same.    

\section{ Summary} To summarize, we have considered a periodic one-dimensional array of two-level emitters with disorder in transition frequencies that are asymmetrically coupled through a waveguide mode   and  have revealed a delicate competition between the conventional Anderson localization and the chiral localization.  {As a result, at moderate values of asymmetry parameter the chirality suppresses the localization, which leads  to} the transition from bulk localized states to edge localized states via  extended states.

We believe that our findings will be important for a rapidly developing field of waveguide-QED, where  chiral interactions and disorder play a critical role. {The modern experimental setups and platforms such as fiber-coupled cold atoms \cite{Mitsch2014, Corzo2016} and superconducting circuits \cite{Guimond2020} have been already used for observing chiral interactions in complex quantum systems. The estimated and experimentally reported range of asymmetry parameter $10^{-3}<\xi<10^{-1}$\cite{Mitsch2014, Corzo2016, Guimond2020, Vetsch2010, Vermersch2016} enables observation of the predicted delocalization effects in realistic systems}.  Moreover, the  extension of the obtained results to multiphoton domain will be of significant interest due to a tremendous progress of theoretical \cite{Poddubny2019,Mahmoodian2018,Mahmoodian2019} and experimental studies in this area \cite{Besedin2021,Prasad2020} { as well as generalization of our approach to non-stationary Floquet-type systems  \cite{Lubatsch2010, Frank2012, Lubatsch2019,Lubatsch2019a}} .    

\section{Acknowledgements}

The authors are thankful to Ivan Iorsh, and Vladimir Yudson for fruitful discussions. 
Numerical and analytical calculations using transmission coefficient  performed by GF, DK and MP, were funded by RSF Grant No. 21-72-10107, and calculations involving eigenstates performed by DK were funded RSF Grant No. 21-72-00096. MP also acknowledges the support by the Priority 2030
Federal Academic Leadership Program, and also from the Foundation for the Advancement of Theoretical Physics and
Mathematics “BASIS.”

\appendix

\section{Dispersion of an infinite periodic chain }
\label{App A}

In this section, we derive the dispersion relation for an infinite ordered chain with an arbitary coupling directionality $\xi$. 
Namely, starting from the following ansatz for the eigenstate wave function $\ket{\varphi} =\sum_{n=-\infty}^{+\infty}e^{iqan}\ket{n}$ (where $\ket{n}$ state corresponds to $n$-th emitter being excited, while all the others are in the ground state) and substituting into the equation:
\begin{align}
\widehat{H}\ket{\varphi} = \widehat{H_0}\ket{\varphi} + \widehat{V}\ket{\varphi} = \hbar\left(\omega - i\frac{\gamma_0}{2}\right)\sum\limits_{\mathclap{{n=-\infty}}}^{+\infty}e^{iqan}\ket{n} +\nonumber \\  \hbar\sum\limits_{\mathrlap{{n,m=-\infty}}}^{+\infty}g_{m,n}
\widehat\sigma_{n}^+\widehat\sigma_{m}^-\sum\limits_{n'=-\infty}^{+\infty}e^{iqan'}\ket{n'}  = E\ket{\varphi},
\end{align}
one can obtain the following dispersion relation:
\begin{align}\label{eq:dispersion}
\Delta\omega(q)= \dfrac{\gamma_0}{2(1+\xi)}\left[ \cot\left(\dfrac{\varphi-qa}2\right)\right. & +\left.\xi \cot\left(\dfrac{\varphi+qa}2\right)\right],
\end{align}
which can also be found in Ref. \cite{BakkensenArxiv2021}.

\begin{figure}[t]
	\includegraphics[width=0.8\columnwidth]{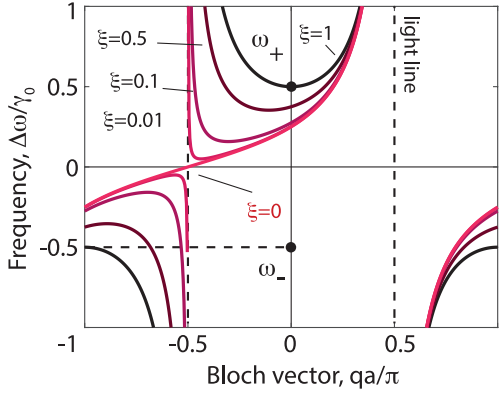}
	\caption{The dispersion of polaritonic modes in a regular array with account for chiral interactions shown for different values of asymmetry parameter $\xi$.}\label{fig:S1}
\end{figure}

The dispersion curves calculated for three characteristic values of the parameter $\xi$ is shown in Fig.~\ref{fig:S1}. The  change of the asymmetry parameter from $\xi=1$ to 0 makes the dispersion nonreciprocal, $\Delta\omega(q)\ne \Delta\omega(-q)$. It also leads to the closure of the band gap, as described by the equation \[E_g=\omega_+-\omega_-=\frac{2\sqrt{\xi}}{1+\xi}\gamma_0\:.\]

\section{Finite regular chain}\label{App B}
Once the  system becomes finite, the non-zero radiative losses appear due to photon escape at the edge of the array, and the eigenfrequencies of the collective states acquire the imaginary parts following Eq.~\eqref{eq:eigenmodes}. The wavefunctions of the eigenstates $\ket{\psi_k}=\sum_n c_{nk}\ket{n}$ can be found analytically both for symmetric and asymmetric coupling  {
by generalizing the results of Ref.~\cite{Voronov2007JLu}:
\begin{align}\label{eq:P2}
c_k&=e^{\rmi q_+(k-N-1)}+r_{\hookleftarrow}e^{\rmi q_-(k-N-1)}\\ 
&\propto 
r_{\hookrightarrow}\e^{\rmi q_+k}+\e^{\rmi q_-k}
,  \nonumber
\end{align}
Here, the wavevectors $q_\pm$ satisfy the dispersion equation 
Eq.~\eqref{eq:dispersion} at the eigenmode frequency $\Omega$ and are chosen in such a way that
$\Im q_+>0$, $\Im q_-<0$. The representation Eq.~\eqref{eq:P2} shows that the eigenmode of a finite array is given by a superposition of corresponding forward- and backward- propagating Bloch waves of the infinite system. The Bloch waves transform into each other  due to the reflection at the internal left and right boundaries of the array with the corresponding reflection coefficients:
\begin{equation}\label{eq:rcoeff}
    r_{\hookrightarrow}=-\frac{1-\e^{\rmi (\varphi-q_+)}}{1-\e^{\rmi (\varphi-q_-)}},\quad r_{\hookleftarrow}=-\frac{1-\e^{\rmi (q_-+\varphi)}}{1-\e^{\rmi (q_++\varphi)}}\:.
\end{equation}
The two representations in Eq.~\eqref{eq:P2} are equivalent to each other because the following identity holds at the eigenmode frequency: 
\begin{equation} \label{eq:determinant}
{r_{\hookleftarrow}(\Omega)r_{\hookrightarrow}(\Omega)\e^{\rmi (q_+-q_-)(N+1)}=1}\:.
\end{equation}
Equation~\eqref{eq:determinant} is a closed-form equation that can be used to find the eigenfrequencies $\Omega$.  It has the same physical meaning as the Fabry-Perot  condition for the eigenmodes of a planar cavity.  The only difference is that the problem is now discrete, and instead of just forward- and backward going photons we consider polaritonic waves.
In practice, however, Eq.~\eqref{eq:determinant} is not easier to solve than the linear eigenproblem Eq.~\eqref{eq:eigenmodes}.
}
 \begin{figure}[t]
	\includegraphics[width=\columnwidth]{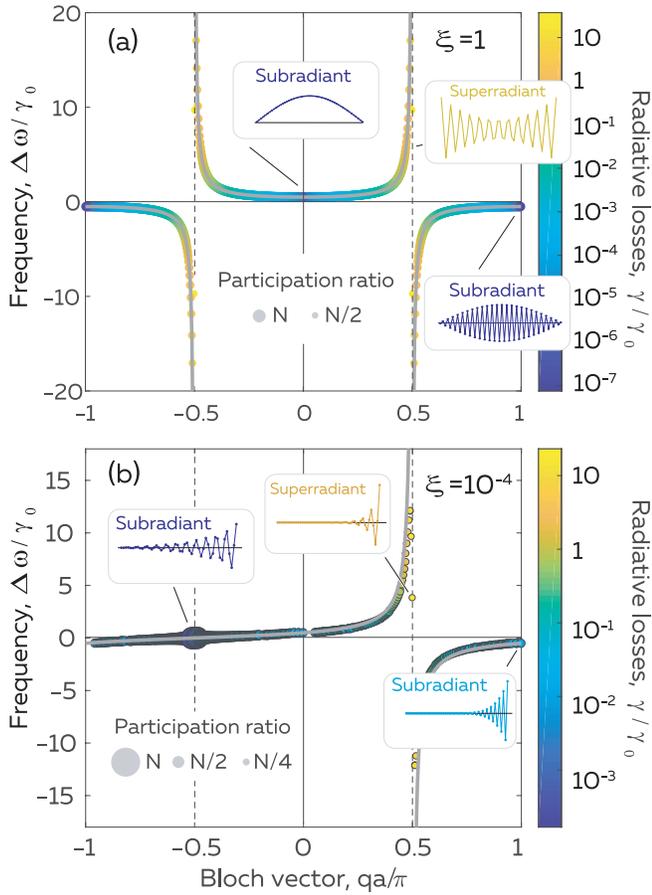}
	\caption{{(a)}  The resonant states of symmetrically  coupled ($\xi=1$) array of $N=400$ quantum emitters separated with $\varphi=\pi/2$. The dispersion of the infinite system is shown with solid grey line. The color of labelling point  denotes the normalized radiation loss rate for each state. The diameter of the labelling point corresponds to normalized  participation ratio. The typical $PR$ values are shown for eye guidance in the inset. Mode profiles  in the insents are plotted for $N=50$ for clearness. (b) The resonant states of chirally coupled array with asymmetry  parameter $\xi=10^{-4}$. The  parameters of computation are the same as in (a).
		\label{fig:2_Disp}}
\end{figure}

We now discuss these eigenmodes in more details in specific cases of symmetric and asymmetric coupling.

 {\it Symmetric coupling.} In the case of symmetric coupling  the  eigenfrequencies  form a circular  structure  \cite{Vladimirova1998} in the complex plane typical  for Toeplitz-type matrices \cite{Movassagh2017}. In order to plot the dispersion of a finite system, one can map the obtained eigenfrequencies of  collective states to the first Brillouin zone of an infinite structure. 

 The eigenfrequencies for an array of  $N=100$ emitters, and the phase parameter $\varphi=\pi/2$ are plotted in Fig.~\ref{fig:2_Disp} for (a) symmetric $\xi=1$, and (b) asymmetric $\xi=10^{-4}$ coupling. They form a discrete set of points on the dispersion line of the infinite system (grey solid line in Fig.~\ref{fig:2_Disp}). The color of  points in the figure denotes the radiative decay rates for each particular state, clearly showing that the states close to the band edge have the smallest decay rate (subradiant), while the states close to the avoided crossing region possess the strongest radiative losses due to the better phase matching with the waveguide mode.  In the insets, we plot the distribution of wavefunction amplitudes  $|c_{nk}|^2$.

\begin{figure}[b]
	\includegraphics[width=\columnwidth]{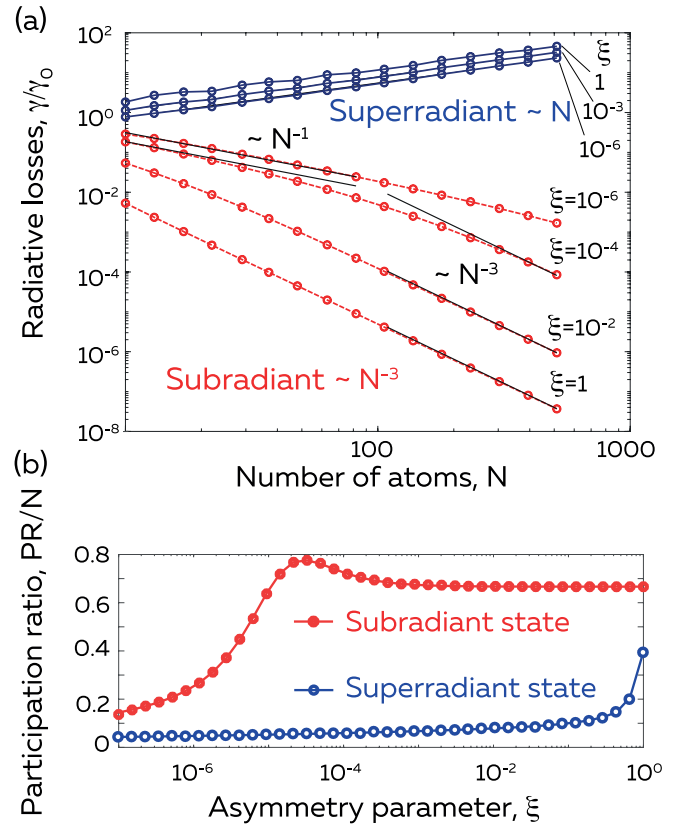}
	\caption{ (a) The radiative losses of the  states with largest (superradiant) and smallest (subradiant) values of $\gamma$ as a function of  number of emitters for different value of asymmetry parameter  $\xi$ and $\varphi=\pi/2$. (b) The participation ratio of the corresponding super- and subradiant states as a function of asymmetry parameter for $N=400$ and $\varphi=\pi/2$.  \label{fig:3_Gamma_PR}}
\end{figure}

 The radiative losses of subradiant and superradiant states  scale with the  size of the system as $\gamma_{sub}\propto N^{-3}$ \cite{Asenjo-Garcia2017,Zhang2020b}, while the emission rates of superradiant states $\gamma_{sup}\propto N$ \cite{Dicke1954, Haakh2016}. The  radiative losses scaling law on $N$ is shown in Fig.~\ref{fig:3_Gamma_PR} (a) for symmetric coupling $\xi=1$.   
 On the other hand, the effective distance to the edge of the array $L_q$ also changes with $\xi$ and gives its contribution to the modified radiation rate. 
 
 Since in the case of the ordered structure, the  eigenmodes of the system are constructed from the Bloch waves, the excitation occupies almost all of the lattice sites $PR\sim N$. We have depicted the normalized participation ratio $PR/N$ for each mode in Fig.~\ref{fig:2_Disp} (a)  with the diameter of the circle labelling the PR value for each state. One can see, that the superradiant states have the smallest PR, while the subradiant states, on the contrary, are the most extended ones with $PR\approx N$. Moreover, all of the states in the ordered array scale linearly with the system size, so $PR\propto N$, which is a sign of their truly extended nature.  
 
One also needs to mention a special case  of $\varphi=0$, which corresponds to a discrete Bardin-Cooper-Schrieffer model \cite{Modak2016,Celardo2016} and is proposed for the description of superconducting states in lattice models. In this case, there appear $N-1$ degenerate states with zero radiative rate and one non-degenerate state, which has superradiant character with $\gamma_N=N\gamma_0$ and constant mode profile with in-phase amplitudes $\ket{\psi}=1/\sqrt{N}\sum_n\ket{n}$.

{\it Asymmetric coupling.} Once the strongly asymmetric coupling is introduced for a finite system,  discrete resonant states  follow the dispersion behavior of an infinite structure as shown with a solid grey line in Fig.~\ref{fig:2_Disp} (b). One can see that the avoided crossing at $qa=-\varphi$  vanishes for asymmetric coupling, and the resonant states close to this point posses  the lowest radiative losses.  Interestingly,  the radiative losses of  subradiant states have different scaling with $N$ comparing to a symmetric coupling case as one can see  from Fig.~\ref{fig:3_Gamma_PR}(a), where the radiative losses are plotted in double logarithmic scale as functions of the emitter number $N$ for various  values of the asymmetry parameter $\xi$. It is worth noting that for small asymmetry parameter values the decay rate of subradiant states scales as $\gamma_{sub}\sim N^{-1}$ for $N\lesssim N^*$, while for larger $N\gtrsim N^*$ the scaling modifies to $\gamma_{sub}\sim N^{-3}$.  As can be seen from Fig.~\ref{fig:2_Disp} (b),  subradiant states appear close to the light line $qa=-\varphi$, where the band gap shrinks with the decrease of $\xi$ as well as the region of the flat band, where the group velocity tends to zero. The switching between the linear dispersion regime to the flat band regime in the vicinity of the light line $qa=-\varphi$ provides the change in the radiative rate behaviour. Indeed, the  radiative decay rate of a state can be estimated as $\gamma(q)\sim v_g(q)/L_q$,  where $L_q$ is the  characteristic distance from the mode center to the structure edge, and $v_g$ is the group velocity.  The group velocity rapidly changes in the vicinity of the light line from $v_g=q \gamma_0/4$ to $v_g\to 0$ as $|qa-\varphi|\sim1\sqrt{\xi}$, which provides the observed change in the scaling of the radiative losses  and gives the estimation for $N^*\sim1/\sqrt{\xi}$.

The second factor which results in decrease of the radiative rate of chirally coupled systems compared to symmetric coupling is related to change of the PR which also corresponds to characteristic length $L_q$, i.e. the smaller is $PR$ the faster the states escape through the edge by radiation. The  dependence of the PR on the asymmetry parameter is shown in Fig.~\ref{fig:3_Gamma_PR}(b). One can see that $PR$ for both sub- and superradiant states decreases  with $\xi$, which  increases the radiative rate for strongly asymmetric coupling.   
 

Subradiant states have the largest $PR$ values close to $N$, therefore, the excitation occupies most of the array sites as shown by the label diameter in   Fig.~\ref{fig:2_Disp}(b). However, now the modes become  localized at the edge of the chain as it is shown in the insets of Fig.~\ref{fig:2_Disp}(b). If $\xi$ becomes small enough, the excitation in the system is concentrated at the right side of the chain as has been discussed in the main text.

\section{Basic formulas for transfer matrices}
\label{AppB}
The forward transmission coefficient $T_N\equiv |t_N^\rightarrow|^2$ in Eq.~\eqref{eq:Lloc0} can also be calculated numerically. This allows us to implement an independent calculation of the localization length, not relying on the evaluation of the eigenstates. To this end we use the transfer matrix method. Starting from the relation between the fields to the left and
right of the atom:
\begin{equation}
\left(\begin{array}{c}
E_{\overrightarrow{R}} \\
E_{\overleftarrow{R}}
\end{array}\right)=M_{\text {atom }}\left(\begin{array}{c}
E_{\overrightarrow{L}} \\
E_{\overleftarrow{L}}
\end{array}\right)
\end{equation}
one can define transfer matrix through a two-level atom $M_{atom}$ \cite{Corzo2016}:
\begin{equation}
    M_{\rm atom}=\frac{1}{t_\leftarrow}\begin{pmatrix}
t_{\rightarrow}t_{\leftarrow}-r^2&r\\-r&1
\end{pmatrix}
\end{equation}
where $r$ and $t_{\rightarrow/\leftarrow}$ are reflection and forward/backward transmission coefficients of a single atom, respectively, given by \cite{Lodahl2017}:
\begin{align}
r&=\frac{\rmi \sqrt{\gamma_L\gamma_R}}{\omega_0-\omega-\rmi \gamma_0/2},
\nonumber\\
t_{\rightarrow/\leftarrow}&=1+\frac{\rmi \gamma_{R/L}}{\omega_0-\omega-\rmi \gamma_0/2}\:.
\end{align}
With transfer matrix for a free part of the waveguide $M_d$ being equal to:
\begin{equation}
M_{d}=\left(\begin{array}{cc}
\mathrm{e}^{\mathrm{i} \omega d / c} & 0 \\
0 & \mathrm{e}^{-\mathrm{i} \omega d / c}
\end{array}\right),
\end{equation}
we proceed to the total transfer matrix through an array of $N$ atoms periodically placed with the distance
$d$ as follows: 
\begin{equation}
    M_N = (M_dM_{\rm atom})^N,
\end{equation}
and find  reflection and transmission coefficients for the light incident from left 
as:
\begin{equation}
    r_{N}^{\leftarrow}=-\frac{\left[M_{N}\right]_{2,1}}{\left[M_{N}\right]_{2,2}}, \quad t_{N}^{\rightarrow}=\frac{\operatorname{det} M_{N}}{\left[M_{N}\right]_{2,2}}.
\end{equation}

\section{Disorder in a perfectly unidirectional system}
\label{appendix:chiral_disorder}
The effective Hamiltonian for a  regular $1$D array of atoms that are unidirectionally coupled through a guided mode and experience a small disorder in transition frequencies can be formally expressed as
\begin{gather}
H_{\text{eff}} = 
\begin{pmatrix}
D_1 & 0 & 0 & \ldots & 0 \\
g e^{i \phi} & D_2 & 0 & \ldots & 0 \\
g e^{i 2 \phi} & g e^{i \phi} & D_3 & \ldots & 0 \\
\ldots & \ldots & \ldots & \ldots & \ldots \\
g e^{i (N-1) \phi} & g e^{i (N-2) \phi} & g e^{i (N-3) \phi} & \ldots & D_N
\end{pmatrix},
\end{gather}
where $g = -i\hbar \gamma_0$, $D_k = \hbar \left( \Delta \omega_k - i  \dfrac{\gamma_0}{2} \right)$, and $\phi = k_0 d$. Owing to a disorder, the degeneracy of eigenstates, appearing as a result of interaction unidirectionality, is completely lifted. In this case we have a full set of eigenvectors $\mathbf{v^{(k)}}$ with the corresponding eigenvalues being equal to $\lambda_k = D_k$, and the latter simply comes from the $H_{\text{eff}}$ matrix being a lower triangular one. One can explicitly find the $j^{\text{th}}$ component of eigenvector $\mathbf{v^{(k)}}$  to be equal to:
\begin{multline}
    v^{(k)}_j = A_k \Big( \delta_{j, N} + \left( 1 - \delta_{j,N} \right) H\left[j - k\right] \\ \prod\limits_{m = j+1}^{N} \frac{\left( D_k - D_m \right) }{ \left( g + D_k - D_{m-1} \right) } e^{-i(N-j) \phi} \Big),
\end{multline}
where $A_k$ is the normalization constant, $H\left[j - k\right]$ is a discrete Heaviside function being zero for $k>j$, and $1$ otherwise. Even though the above formula is slightly cumbersome, it is much easier to understand if one considers the transformation matrix to the corresponding eigenspace $S = \left( \mathbf{v^{(1)}}, \: \mathbf{v^{(2)}}, \: \ldots, \: \mathbf{v^{(N)}} \right)$:
\begin{widetext}
\begin{gather}
     S = \begin{pmatrix}
    A_1 \frac{(D_1 - D_2) \ldots (D_1 - D_N) e^{-i(N-1)\phi}}{(g + D_1 - D_1) \ldots (g + D_1 - D_{N-1}) }  & 0 & \ldots & 0 & 0 \\
    A_1 \frac{(D_1 - D_3) \ldots (D_1 - D_N) e^{-i(N-2)\phi}}{(g + D_1 - D_2) \ldots (g + D_1 - D_{N-1}) }  & A_2 \frac{(D_2 - D_3) \ldots (D_2 - D_N) e^{-i(N-2)\phi}}{(g + D_2 - D_2) \ldots (g + D_2 - D_{N-1}) }  & \ldots & 0 & 0 \\
    \ldots & \ldots & \ldots & \ldots & \ldots \\
    A_1 \frac{(D_1 - D_{N}) e^{-i \phi}}{(g + D_1 - D_{N-1}) }  & A_2 \frac{(D_2 - D_{N}) e^{-i \phi}}{(g + D_2 - D_{N-1}) }  & \ldots & A_{N-1} \frac{(D_{N-1} - D_{N}) e^{-i \phi}}{(g + D_{N-1} - D_{N-1}) }  & 0 \\
    A_1 & A_2 & \ldots & A_{N-1} & A_N
    \end{pmatrix}.
\end{gather}
\end{widetext}

\begin{figure}[t]
	\includegraphics[width=0.9\columnwidth]{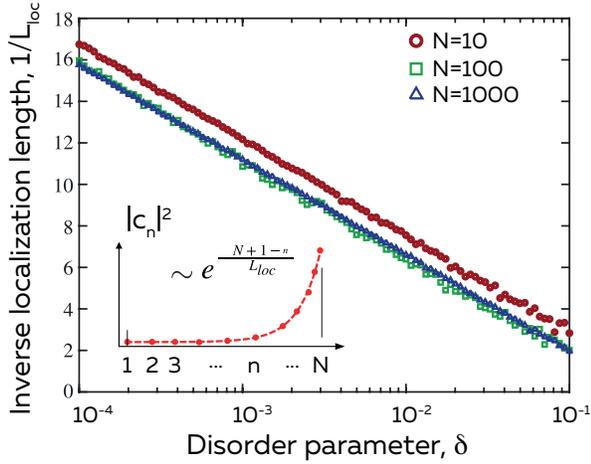}
	\caption{ Dimensionless inverse localization length $L_{\text{loc}}^{-1}$ versus disorder parameter $\delta$ for different number of atoms in a chain of $N$ unidirectionally coupled atoms $\xi=0$. It has been extracted by fitting an exponential function to the probabilities of few atoms closest to the right edge of the chain being excited, as described by the inset.} 
	\label{fig:fig_apx_1}
\end{figure}

As seen from the above, the matrix $S$ is also lower triangular, and the $N^{\text{th}}$  eigenstate corresponds to the last atom being excited solely, similarly to the case of absent disorder, which is a perfectly localized state with the corresponding participation ratio being $PR (\mathbf{v^{(N)}}) = 1$. A simple inspection tells that, obviously, all other states $k \ne N$ have $PR > 1$, but, simultaneously, even the state $\mathbf{v^{(1)}}$ with all non-zero components is not a delocalized one due to the fact that $\text{var}\left( \Delta \omega_k \right) = \delta \gamma_0 \le 0.1 \gamma_0$ in our case. This is what is indicated in Fig.~\ref{fig:fig_apx_1}, where the dimensionless inverse localization length $L_{\text{loc}}^{-1}$ (an effective number of excited atoms) is plotted against the disorder strength $\delta$ for different number of atoms in a chain. As seen, $L_{\text{loc}}^{-1}$ monotonically decreases with the disorder $\delta$ following a logarithmical dependence almost perfectly. Moreover, even for the largest disorder parameter considered $\delta = 0.1$,  $L_{\text{loc}}^{-1} > 1$, which means that localization length is smaller than unity, hence we conclude that the most delocalized state in terms of $PR$ is, strictly speaking, a localized one. We can conclude that for a perfectly chiral case, the introduction of disorder into atomic transition frequencies {does not lead to localization-delocalization transition for the considered range of a disorder parameter $\delta$}.

\bibliography{Biblio}

\end{document}